\documentclass[onecolumn,notitlepage]{article}
\usepackage{graphicx}
\usepackage{amsmath}
\usepackage{epsfig}

\begin{document}

\title{Pattern Excitation-Based Processing:\ The Music of The Brain}
\author{Lev Koyrakh\thanks{%
e-mail:\ levkoyrakh@physics.spa.umn.edu} \\
University of Minnesota}
\date{October 14, 2003}
\maketitle

\begin{abstract}
An approach to information processing based on the excitation of patterns of
activity by non-linear active resonators in response to their input patterns
is proposed. Arguments are presented to show that any computation performed
by a conventional Turing machine-based computer, called T-machine in this
paper, could also be performed by the pattern excitation-based machine,
which will be called P-machine. A realization of this processing scheme by
neural networks is discussed. In this realization, the role of the
resonators is played by neural pattern excitation networks, which are the
neural circuits capable of exciting different spatio-temporal patterns of
activity in response to different inputs. Learning in the neural pattern
excitation networks is also considered. It is shown that there is a duality
between pattern excitation and pattern recognition neural networks, which
allows to create new pattern excitation modes corresponding to recognizable
input patterns, based on Hebbian learning rules. Hierarchically organized,
such networks can produce complex behavior. Animal behavior, human language
and thought are treated as examples produced by such networks.
\end{abstract}

\tableofcontents

\section{Introduction}

The goal of this paper is to present a theory capable of explaining the
remarkable information processing abilities and complex behavior of
biological neural networks found in living organisms, particularly in their
brains. To accomplish this goal an information processing (computing)
paradigm is discussed which is based on the view of the computational
process as being an excitation of a system in response to its inputs,
similar to the excitation of a particular mode of a string in response to a
particular pattern of its initial velocities and displacements. The proposed
processing paradigm is different from the one based on Turing machines, and
because it is based on excitation of output patterns by some kind of \textit{%
resonators} it will be called \textit{pattern excitation - }based. A
generalized machine realizing such processing will be called \textit{%
P-machine}. One of the statements of this paper is that any computation
achievable with the use of conventional computers based on Turing machines
(called here T-machines for brevity) is also achievable by P-machines.

The term \textit{resonator} is used in this paper to denote an excitable
object of some nature which is capable of producing different modes in
response to different input patterns. It is not required to be linear, and
in fact, non-linearity could be essential for the ability to add new
excitation modes to the set of exiting ones. The excitation modes of the
resonators discussed here could be changed. The process of \textit{tuning}
of the resonators includes adjustments to the excitation patterns of the
existing modes and adding new modes, and can be called \textit{learning}.
Tuning a resonator in a P-machine is similar to re-programming a
T-machine-based computer. The property of the resonators of being \textit{%
active} means that they can draw energy from sources different from their
inputs.

There could be, and in fact exist in the real world, many realizations of
P-machines with different levels of complexity, musical instruments being
among the ''simplest'' of them. The realization of P-machines which is of
the main interest in this paper is given by neural networks.

The brain's main information processing cells are neurons, which make
numerous synaptic connections with other neurons forming what is known as
neural networks. The remarkable abilities of neural networks in respect to
pattern recognition are being intensively studied (see books and reviews 
\cite{dayan},\cite{compbrain},\cite{haykin},\cite{theory}). It has also been
noticed that neural networks are, in principle, capable of performing any
computation that a conventional computer can do.

The main power of neural networks comes from the parallel processing of
information by very many neurons. But how exactly are neurons organized for
the information processing? The well studied neural networks performing
pattern recognition function could be considered as passive processing
elements. The active elements are given by the pattern excitation networks
formed by neurons. The pattern excitation circuits considered here are
capable of creating spatio-temporal patterns of activity and are considered
to be the network's main components which code and generate all responses of
the network to all potential inputs.

By stating that neurons are organized in the two primary circuit types, one
effectively performs what is known in physics as ''renormalization'', which
in this case amounts to replacing fine degrees of freedom represented by
neurons with the more coarse grained ones represented by the pattern
excitation and recognition circuits. These coarse grained degrees of freedom
provide the representation of the nervous system which is more adequate for
describing how it works, including behavior and cognition. This transition
to the coarse grained picture efficiently reduces the number of degrees of
freedom by few orders of magnitude and makes the system much easier to
study. Individual neurons, however, still have a role to play: they are
wires connecting different parts of the neural P-machine.

One can think of the pattern excitation networks as adjustable, tunable
resonators made out of individual neurons, capable of learning and
reproducing a number of excitation modes. Various modes of these resonators
could be excited by various patterns of stimuli (inputs).\ The process of
tuning the resonators is equivalent to learning by the neural network.

It will be shown below that training of an excitation pattern network could
be accomplished by the corresponding pattern recognition network in the
supervised learning mode, assuming Hebbian learning in the networks. This
leads to the concept of dual pattern recognition - pattern excitation
networks, which recognize inputs and generate responses to them.

Hierarchically organized pattern excitation networks can produce very
complex behavior and could be responsible for human language and cognition.
Some approaches on how that could be accomplished will be considered in
chapter \ref{ch examples}.

In the computer science language, excitation patterns play the role of
different subroutines or procedures called by the main program for
processing of various inputs. The main program itself, which often takes the
form of an unconditional loop, could be compared to the set of excitation
patterns, which control and organize all other excitation pattern circuits.
Perhaps even more adequate would be comparison of pattern processing with
the object oriented approach in computer science, in which pattern
recognition and excitation networks would be direct analogs of objects.

The term\textit{\ neural networks} is usually used to describe artificial
neural systems. For terminological simplicity in this paper this term will
be used for description of both artificial and natural neural systems. Also
the words \textit{network} and \textit{circuit} will be used
interchangeably. The pattern recognition networks will be denoted as PRN,
and pattern excitation networks will be denoted as PEN. Since the only
realization of P-machines considered in this paper is based on neural
networks, the discussion of P-machines will be made mostly in terms of their
neural network-based realization.

The work on this paper was largely stimulated by observations that neural
networks and brain exhibits some properties of systems studied in the
condenced matter physics, particularly by the works \cite{theory},\cite
{leasteffort},\cite{chess1},\cite{chess2}.

\section{P-machines}

\begin{figure}[tbh]
\centering
\epsfig{figure=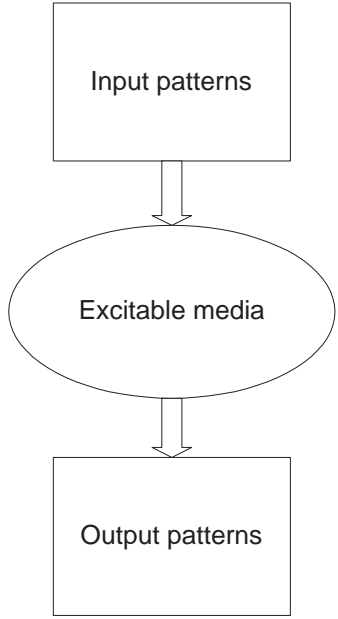,width=1.75in}
\caption{P-machine takes inputs and finds patterns in them which in turn
excite correponding pattern-generating networks. The generated patterns
serve as the P-machine outputs.}
\label{p-machine}
\end{figure}

A P-machine is a mathematical abstraction of the pattern-based information
processing. A generalized P-machine is a set of resonators. Particular
nature of these resonators is not important here, it could be strings,
membranes, resonance cavities of any sort, acoustical, electromagnetic, or
other. It is illustrated on Fig. \ref{p-machine} where the set of resonators
is called \textit{excitable media}. The resonators considered here could be 
\textit{active }and\textit{\ non-linear}, meaning that they could be capable
of amplifying particular input patterns, or exciting output patterns not
necessarily resembling the input patterns. Each resonator from the set could
be excited by patterns of its initial distortion of the corresponding
nature. The exact mode of the excitation is the function of the input
pattern. This excitation could be thought of as the result of the
''computational process'' coded by the resonator. Tuning the resonator to a
different set of excitation modes could be thought of as re-programming of
the resonator. If the modes of the resonators could be changed by the input
patterns, then memory could be created in such systems. One of the
statements of this paper is that any computation possible by the Turing
machine is also possible by a set of tunable resonators, and vice versa. \ A
note is in place, however, that the mapping of one machine into another is
not necessarily direct. For this mapping, the work of the Turing machine has
to be represented as a set of recognizable patterns capable of exciting
other patterns in the corresponding P-machine, so each excitation could
correspond to a complex program run on the Turing machine. Vice versa, a
single instruction of a Turing machine cold be translated in a potentially
very complex set of excitation patterns of the P-machine.

Any combination of P-machines is again a P-machine.

Depending on their construction, some P-machines can learn to recognize new
input patterns and excite new output patterns. This learning on the one hand
is the intrinsic property of the P-machine, and on the other, could \
require a feedback from its environment. Generally speaking, a P-machine
takes its inputs from its environment, and its ouptut could be considered as
a change in the environment. This closes the fedback loop, since the next
input to the P-machine will be provided by the changed environment.
Therefore, continuous operation of a P-machine can result in significant
changes to its environment.

To formally describe a P-machine, let us introduce the following notation:

\bigskip

$t$ \ \ \ - time

$I$ \ \ \ - \ the input field (example:\ a set of neurons providing the
input to a circuit)

$O$ \ \ - the output field (example: a set of neurons taking the output from
a circuit),

$p_{I}$ \ - \ the set of patterns recognized in the input field

$p_{E}$ \ - the set of excitation patterns produced by the excitable
resonators

$p_{O}$ \ - the set of output patterns of the P-machine

The computation process by the P-machine could be described by the following
set of expressions ($f$ stands for some functions): 
\begin{equation}
p_{O}(t)=f_{OI}(p_{E}(t)),  \label{po}
\end{equation}
which means that the output patterns are results of the pattern excitation
activity of the machine. In turn, 
\begin{equation}
p_{E}(t)=f_{EI}(p_{I}(t)),  \label{pe}
\end{equation}
meaning that the excitation patterns are generated by the machine in
response to recognized patterns in the input field, and 
\begin{equation}
p_{I}(t)=f_{II}(I(t)).  \label{pi}
\end{equation}
These equations must be supplemented by the machine learning rules
describing how the sets of the recognized and excited patterns are changed: 
\begin{eqnarray}
p_{I} &=&L_{I}(I,p_{E}) \\
p_{E} &=&L_{E}(p_{I},p_{E}).  \label{pel}
\end{eqnarray}
For these equations to be useful in particular applications, all functions $%
f $ \ and the learning rules $L_{I}$ and $L_{E}$ must be specified.

The correspondence between the pattern processing based machines and Turing
machines can be loosely formulated by the following statement.

P-machines are capable of performing any computation a T-machine can perform
and vice versa.

To prove that any computation possible by a P-machine can also be performed
with a T-machine, one has to write a computer program for a T-machine
simulating the P-machine, and to prove that any computation possible by a
T-machine is also possible by a P-machine, one has to design a P-machine
simulating the T-machine.

This statement says nothing about the efficiency of the maps between the T
and P -machines and resources required for the simulations.

One can think of the patterns excitable by the P-machine as corresponding to
various pieces of a conventional computer program. For example, an
excitation pattern circuit could correspond to a single character,
instruction, a block, a subroutine or a whole very sophisticated program.

\section{Hierarchical structure of patterns and P-machines}

In the MATHEMATICA\ book by Stephen Wolfram \cite{wolfram} there is a
statement that ''\textit{Everything is an expression}''. This paper is
largely based on the axiom that \textit{Everything is a pattern}. An
expression is also a pattern. Everything from a single symbol to a very
abstract theory could be represented as patterns or patterns of patterns.\
Even the lack of patterns could be classified as a \textit{lack of patterns}
pattern. But what precisely is a pattern?

In conventional computer science, patterns are defined as \textit{extended
regular expressions }\cite{flex}. This definition allows to use \textit{%
finite state machines }for finding patterns. But is it possible to define a
pattern without reference to expressions, or it is an\textit{\ elementary
notion }which can not itself be defined? In the later case it is similar to
such objects as \textit{point, straight line} or \textit{plane} in
elementary geometry. One will have to use patterns of some nature to define
what is a \textit{pattern}. The WordWeb dictionary program installed in the
author's computer defines pattern as \textit{A perceptual structure\ }\cite
{wordweb}. The working definition adopted in this paper is that whatever a
P-machine can recognize or produce is a pattern.

Each pattern is made out of elements, which could also be considered
patterns and will be called \textit{subpatterns. }This leads to hierarchical
organization of patterns.\textit{\ }Generally there is more than one way to
identify subpatterns in a pattern, so that one can say that there are
patterns of ways in which subpatterns could be identified.

To illustrate this point, let us imagine music made by a pianist playing on
a piano. The sounds of music represent excitation patterns of the air\
recognizable by our ears. Each sound is a result of a motion excitation
pattern of a particular string of the piano, yet, it is a subpattern in a
passage which is the pattern represented by the sequence of excited strings
and strength and duration of each sound, which is produced as\ the pattern
of strokes of the piano keys, each of which is a subpattern of the last
pattern. Rearrangement of these subpatterns will lead to a differently
sounding passage - the result of different excitation pattern of the piano
keys. In turn, passages are subpatterns in melodies, melodies are
subpatterns of parts of the composition, and so on.

Similar analysis can be applied to speech and language. Sounds are
subpatterns of words, words are subpatterns of small idiomatic expressions,
which are subpatterns of phrases, which are subpatterns of sentences, which
are subpatterns of thoughts. Thus, thoughts are patterns at a certain level
of this hierarchy.

The described hierarchical scheme is not absolute. There could be many
feedback loops in the system. For example, our thoughts, choice of words,
and sounds are influenced by our desires and emotional patterns acting on
all levels of the described hierarchy. Sounds of music affect the emotional
state of the pianist, which affects the way he or she plays.

The hierarchical nature of patterns naturally leads to the notion of \textit{%
hierarchically organized }P-machines. Bigger P-machines could be made by
hierarchically connecting smaller P-machines. In these architecture,
subpatterns of excitation patterns of one level of the hierarchy of
P-machines become input patterns to the next level P-machines.\ Existence of
feedback loops results in excitation patterns produced by different parts of
a P-machine being not completely static and independent, they are influenced
by other excitations present in the system, they could be blocked,
re-excited or amplified, leading to potentially very complex picture.

One possible organization of the neural network realization of the P-machine
corresponding to the described\textit{\ sounds to thoughts} hierarchy will
be considered in chapter \ref{ch example language}

The hierarchical organization described here does not have to be strict. One
can imagine constructions in which outputs from what is considered lower
level into a higher level P-machine, as well as one level of the hierarchy
providing inputs to several levels below.

\section{Assumptions about the properties of neurons and neural networks}

In order to introduce the realization (or \textit{representation} as a
physicist would call it) of P-machines in neural networks, let us first
consider very general properties of neurons assumed in this paper.

In accordance with the contemporary neuroscience, neurons are excitable
cells making many synaptic connections with other neurons, muscles and
whatever other cells they are effecting or innervating. The effects of the
neural coupling on the post-synaptic cells could be excitatory or
inhibitory, depending on the particular connection. In this paper no
specific assumptions are made about particular properties of neurons outside
of what is generally known about them \cite{dayan},\cite{princofneuro},\cite
{exploringbrain},\cite{purves}. For example, it will be assumed that neurons
have a threshold excitation function, but no particular model of that
function will be used in this paper, as well as no specific assumptions
about particular forms of the action potential or firing patterns of
individual neurons will be made. These considerations could be very
important however in discussions of particular excitation patterns, patterns
stability, reliability of the information carried by the action potential
and so on.

It will be also assumed that by the virtue of Hebbian learning, which
involves adjustments of strength of binary interactions between neurons,
neural networks can be trained to produce specific output signals in
response to classes of input signals. This assumption is based on the
intensively studied learning in neural networks (see for example \cite{dayan}%
,\cite{princofneuro},\cite{exploringbrain},\cite{purves}).

A pattern excitation neural circuit can contain a very large number of
neurons, perhaps on the orders of $10^{4}-10^{8}$. When speaking about such
circuits, it will be assumed that all mechanisms necessary to make
excitation patterns stable are present. Among other thing, these mechanisms
could include positive and negative feedback loops, circuits employing
inhibitory synapse and so on. This area currently is not well studied, so
these assumptions will have to be validated by further research.

Since this paper is concerned with the presentation of a rather general
framework, the assumptions made are very general. For specific applications
of ideas presented here, many specific assumptions about neurons and neuron
networks will have to be made.

\section{Nervous systems as P-machines}

This section outlines the main points of the theory that views nervous
systems as P-machines.

\begin{enumerate}
\item  Neurons are assembled into two primary types of circuits: pattern
recognition and pattern excitation ones. There is no sharp division between
these two types of circuits (pattern recognition and pattern excitation) and
both functions could be represented by the same circuit.

\item  Pattern excitation circuits are capable of learning and producing
specific spatio-temporal patterns of excitations in response to various
patterns of stimuli. There could be excitable circuits with different
characteristic time and spatial scales, producing a wide range of different
excitation patterns.

\item  Learning of an excitation pattern by a network is in a number of ways
similar to learning in the pattern recognition networks (for later see for
example, \cite{dayan},\cite{compbrain}). In the approach considered in this
paper, the learning of excitation patterns is accomplished through a
supervised training of the PEN by a PRN, and could be based on Hebbian
learning (see chapter \ref{ch learning}). The pattern generated by the PEN
will be called dual to the input pattern recognized by the PRN.

\item  Potentially a very large number of neural P-machines could be working
in parallel, simultaneously processing\ various parts of patterns.

\item  Pattern excitation neural networks are hierarchically organized,
which allows them to produce complex behavior, such as observed in living
organisms. In the hierarchial organization, parts of excitation patterns (%
\textit{sub-patterns)} produce streams of inputs which excite patterns in
the lower level networks. Sub-patterns of a single pattern of a higher level
PEN could produce sequence of inputs for a single lower level PEN or
simultaneously excite multiple PENs.
\end{enumerate}

\section{Learning excitation patterns:\ recognition - excitation duality 
\label{ch learning}}

Let us now discuss some ways in which the training of new excitation modes
(patterns) could be accomplished.

\begin{figure}[tbh]
\centering
\epsfig{figure=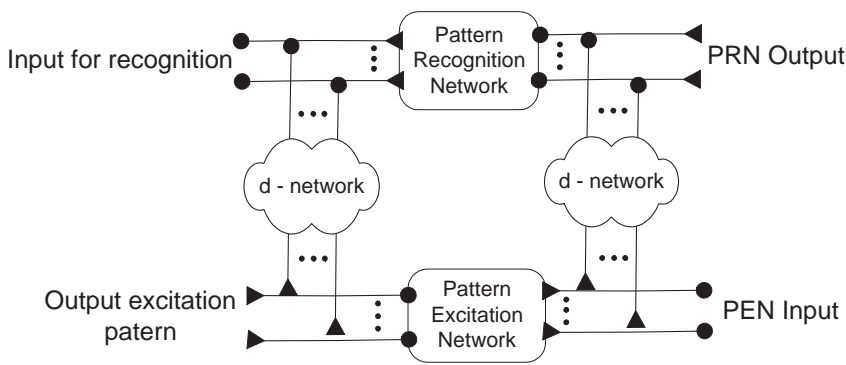,width=5.25in}
\caption{Basic learning in the pattern excitation networks is accomplished
by connecting the input of PEN with the output of PRN\ and the output of
PEN\ with the input of PRN. These connections go thought (delay) d -
networks, so that the pattern recognized by the PRN could serve as the
training input for the PEN, and the input pattern of PRN could serve as the
output training set for the PEN. The d-networks could play roles beyond the
basic delays functions. In fact, they can modify the input to the PRN signal
so that the desired reaction by the PEN is actually trained. These modifying
functions could be selected by the network hosts using various mechanisms,
including evolution through natural selection. Neuron cell bodies are drawn
as circles, axon terminals - as triangles.}
\label{basic-learning}
\end{figure}

Let us connect input of the PRN\ to the output of the PEN, and the output of
the PRN to the input of the PEN using neurons sending axons from the PRN to
PEN, as shown on Fig.\ref{basic-learning} Then when there is an input
pattern of the PRN which it can recognize, the output of the PRN will play
the role of the pattern presented and the input pattern will play the role
of the output shown to the network (PEN)\ under the supervised training.
Because PRN\ output follows its input, delay d-networks are included on Fig. 
\ref{basic-learning} to insure satisfaction of causality conditions in the
training process. If one now assumes that the Hebbian learning is taking
place in the PEN, the PEN will learn an excitation pattern which will
resemble the input. This PEN will be complementary to the PRN\ in the sense
that the input to the PRN\ becomes the prototype of the excitation pattern
of the PEN, while the output of the PRN\ becomes the input to the PEN
triggering the excitation mode producing the pattern. The direction of
connections between the PRN and PEN and,correspondingly the training process
could also be reversed, so that the PEN will train the PRN, provided that
necessary delay circuits are also included.

Another purpose of the d-networks is to alter the input and output of the
PRN which are being delivered for the PEN\ training. By altering the
patterns used in training the PEN, it could be taught a wide range of
responses to various stimuli. It has to be noted that these patterns could
actually be different from the ones experienced by the PRN. The neurons
connecting the two networks could themselves be parts of or be modulated by
different other networks, resulting in the huge variety of possible
excitable patterns stored in the PEN.

\begin{figure}[tbh]
\centering
\epsfig{figure=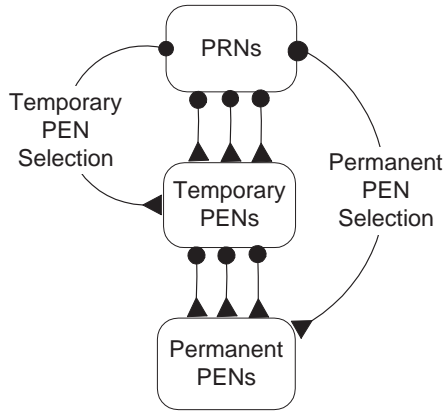, width=3.3in}
\caption{Different variations of learning of excitation patterns could be
considered. In this drawing a temporary PEN\ is used as a scratch memory.
The more a particular pattern is recognized by the PRN, the more permanent
PEN\ learns the corresponding (dual) excitation pattern. In the mean time,
responses are generated by the fast learning Temporary PEN. Selection lines
can contain many neurons, only one neuron is drawn in each line for brevity.}
\label{tpens}
\end{figure}

One can imagine various scenarios in which the described learning mechanism
could be actually implemented in a neural network. For example, a PRN can be
training the formation temporary pattenrs in the PENs serving as a ''scratch
memory'', which in turn could drive the formation of permanent patterns. The
permanent PENs would be strengthened with the use and the PRN\ would also
direct which permanent PEN is trained by the scratch memory at any moment of
time. This approach is illustrated on Fig.\ref{tpens} \ Here new excitation
modes for related recognized patterns could be created in the same pattern
excitation network.

Let us briefly touch upon the relationship between learning and evolution.
Many variations of neural networks organization and learning processes are
possible within the proposed framework.\ For example, d-networks as well as
specific reactions of the network to the same outside input could be
different, though the participating networks could be trained through the
same basic mechanism. These specific reactions and PRN, PEN and d-networks
could be subjects of selection through the evolution of organisms. There is
also a possibility that the PENs are ''pre-trained'' by evolution to the
point when little additional training by the PRNs is needed to adjust the
function of the PENs to specific environments.

The fact that the pattern recognition networks train pattern excitation
networks, which generate reactions of the system to the external inputs,
leads to the picture in which both types of networks are complementary to
each other. The close relationship between them illustrated on Fig. \ref
{basic-learning}, in which the excitation of the PEN are essentially
(modified) inputs to the PRN, allows one to consider the PRN and the PEN as
being dual. Structures like the one shown on Fig. \ref{basic-learning} could
be used as building blocks for more complex networks.\ Hierarchically
organized, such complex networks could be able to perform many functions
which are found in living organisms, including language and cognition.

The folowing remarks are in place.

Though pattern excitation networks are generally capable of learning their
patterns, they do not have to learn anything to work. There could exist PRNs
which just generate patterns, in living organisms this would correspond to
emotions or desires generated without originally learning any patterns. This
would be similar to work of cardiac pacemaker cells, which do not need to be
excited by an action potential passing by. Existence of such ''pacemaker''
networks can also be viewed as the manifestation of a different type of
learning, which could be associated with biological evolution.

The distinction between pattern recognition and pattern excitation networks
is not precise. One can think of the PENs as being PRNs with outputs
described by complex patterns. The situation is somewhat similar to the
distinction between a program and data in conventional computers, where
everything could be treated as data, but one can define the program as being
the part of data which is supplied by the programmer, and the ''true'' data
being all the rest. In the case of neural networks the classification
appears to be trickier and goes by the function of the network in the neural
system. In this case the recognizable by PRNs input patterns can be used to
train excitable networks to respond to different inputs. In fact the
response could manifest itself as very complex behavior, much more complex
than the input that excited it.

\section{Examples \label{ch examples}}

\subsection{Excitable neural circuits}

\begin{figure}[tbh]
\centering
\epsfig{figure=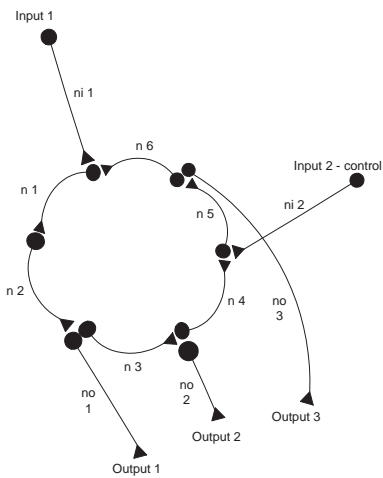, width=2.5in}
\caption{A simple circuit capable of storing and reproducing\ excitable
patterns on its outputs. In this drawing, the excitation patterns could be
triggered by $Input1$ making excitatory synapse with the neuron $n1$, and
modified or blocked by $Input2$ making inhibitory synapse with $n6$.}
\label{exciton}
\end{figure}

A simple example of an excitable neural circuit is given in the Fig. \ref
{exciton}. Detailed analysis of the excitation properties of neural
networks is presented in \cite{ermentrout}. Let me note, that the pattern
excitation networks under consideration in this paper are not necessarily
small, the circuit discussed in this section is chosen as an illustration
for its simplicity.\ Let us assume that the excitatory stimulation of the
neuron cell body (drawn as a circle) will generate an action potential
propagating along its axon, followed by a short refractory period. A stable
excitation traveling counterclockwise could be generated in the circuit by a
pulse stimulus from the input neuron $ni1$, assuming the excitatory synapse
between $ni1$ and $\ n1$.

The excitation wave traveling in the neural circuit will generate a certain
excitation pattern on the output neurons $no1$ - $no3$. The other input
neuron $ni2$ could make either excitatory or inhibitory synapse with some
neuron, $n5$ in this example. This input neuron ($ni2$) has the ability to
alter the excitation pattern of the output neurons or even completely block
the excitation. Despite of the simplicity of the model, it illustrates the
main features of other excitable neural circuits. One can think of the
inputs $ni1$ and $n2$ as providing the input pattern and the outputs $no1$ - 
$no3$ as providing the output pattern which is created by the various
excitation modes (patterns) of the circuit. Different spatio-temporal
combinations of the input signals will result in different output patterns
of activity on $no1$ - $no3.$

More complex circuits are capable of storing and reproduction of more
complex excitation patterns. They could also be controlled in the more
precise and sophisticated way than the simple circuit shown in the above
example.

It should be noted that the circuit considered in this example is not being
treated as a mere memory element, but a computational unit analogous to a
subroutine in an computer program. Every time the circuit is excited, a
specific ''computation'' is performed. The result of this ''computation''\
is coded in the excitation pattern output of the circuit.

Pattern excitation neural networks could be hierarchically organized to
produce complex behavior. As is was already mentioned, each PEN could be
capable of storing and reproducing potentially large number of patterns,
which could be consequently excited by a stream of inputs to the PEN
consisting of subpatterns of patterns excited by a higher level network.
Also, in hierarchically organized PENs, a single excitation of the higher
level network can supply streams of inputs to potentially large number of
lower levels networks.

\begin{figure}[tbh]
\centering
\epsfig{figure=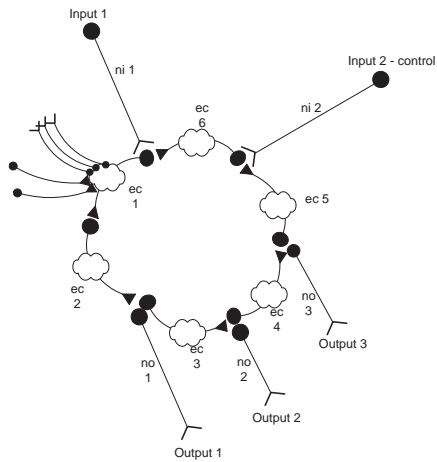, width=2.5in}
\caption{An example of nesting hierarchical neural circuit topology in which
some neurons of \ \ \ \ \ the higher level PEN are replaced by whole next
level PENs. Excitation circuits $ec\,1-ec\,6$ could be also modulated and
excited by additional inputs, and have outputs conducting their excitation
patterns to other circuits (shown for $ec1$ only).}
\label{nesting neurons}
\end{figure}


Various topologies are possible in concrete realizations such as nesting
(Fig. \ref{nesting neurons}), tree and other. Exactly which topologies are
realized and how the network hierarchies are formed and trained is the
subject of further research.

\subsection{Language and thought \label{ch example language}}

The approach to the language and thought processing with P-machines, which
includes recognition and generation of speech and thought, has to be
developed based on the corresponding pattern hierarchy consisting of
patterns of sounds, words, sentences and thoughts. Let us first trace the
thinking and speaking process from the ''bottom-up'', starting with making
sounds and ending with generating thoughts. At each step of the process
there is a set of pattern excitation networks taking inputs from one level
of the hierarchy above it and generating inputs for the networks one level
below.

\begin{enumerate}
\item  Sound pattern excitation networks. Excitation patterns produced by
these networks are passed to muscles making sounds.

\item  Words pattern excitation networks. Excitation patterns of these
networks are passed as inputs to sound excitation networks. Subpatterns of
word excitation patterns trigger excitation of patterns corresponding to
separate sounds. Rearrangements of these subpatterns form patterns
corresponding to different words.

\item  Stable word combinations, idiomatic expressions and grammar pattern
excitation networks. These patterns of combining words into sentences make
up grammar, which in described approach is viewed as a set of patterns in
which individual words are major subpatterns.

\item  Thoughts pattern excitation networks. In this approach thoughts are
defined as patterns at this level of the hierarchy. Subpatterns of these
excitation patterns provide inputs to the lower levels networks described
above.

\item  Emotions and desires pattern excitation networks. These could work as
pacemakers and moderators starting the whole process or regulating its pace\
and mood.
\end{enumerate}

Some of the levels could be subdivided into further sub-hierarchies.

\begin{figure}[th]
\centering
\includegraphics[width=0.63\textwidth]{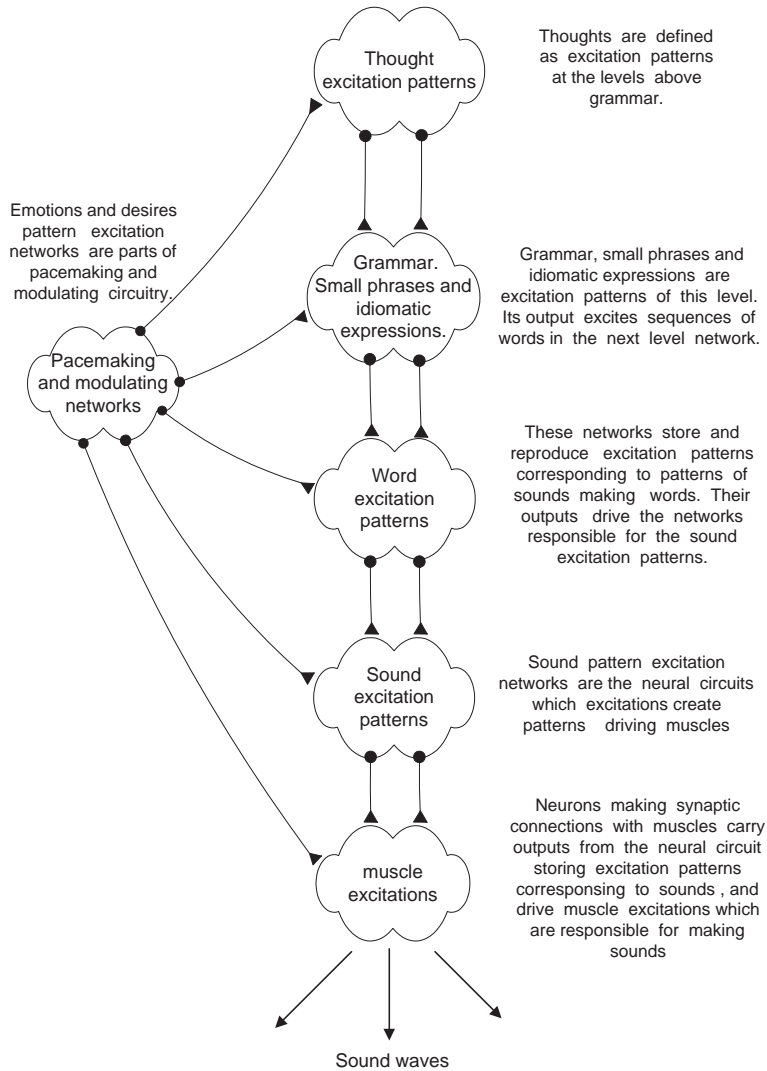} 
\caption{A possible hierarchy of the brain's pattern excitation networks
(PENs) responsible for thought and speech. In this picture, thoughts are the
excitation patterns at the top of the hierarchy. Excitation patterns
produced by thoughts drive the pattern excitation networks representing
grammar, which, in turn, drive the excitable networks responsible for the
more fine structured elements of language all the way down to sounds.
Neurons mediating feedback and connecting distant levels of the hierarchy
are not shown.}
\label{thought hierarchy}
\end{figure}

This approach is illustrated on Fig. \ref{thought hierarchy}. In the
described hierarchy all level are treated as (neural) P-machines, which
could have different complexities.

This hierarchy should be considered as the dominant structure, however other
structures could coexist alongside with it. For example, some subpatterns of
lower level excitations could be fed back to higher level networks, while
the later could provide inputs to more than the one levels below, which
would lead to ''context sensitive'' speech, and thoughts being directed by
the choice of words, or sound of speech.

\begin{figure}[t]
\centering
\epsfig{figure=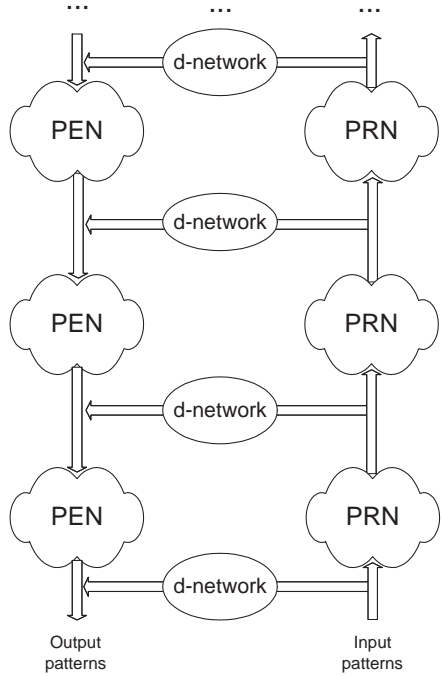, width=2.3in}
\caption{A P-machine capable to recognize its inputs and generate output
patterns in response. It consists of pattern excitation (PEN) and pattern
recognition (PRN) network hierarchies running in parallel. d-networks allow
PENs to be trained by the corresponding PRNs. On this diagram neurons are
drawn as arrows showing the direction of the action potential propagation.
Each level of the hierarchy recognizes by its PRN and excites by its PEN
patterns of the same level in the pattern hierarchy.}
\label{ladder}
\end{figure}

In addition to the excitation networks, a parallel hierarchy of recognition
networks can be drawn, as shown on Fig. \ref{ladder}. The PRNs shown on this
figure are connected to their dual PENs, so that PENs could be trained. The
resulting P-machine is capable to recognize its inputs and generate
responses to them as output patterns.

Let me note that the recognition process by the hierarchy of PRNs based on
the hierarchy of patterns is different from translating computer languages,
which is based on ''parsing an expression'' \cite{languagepragmas},\cite
{flex},\cite{bison}.

Talking about the human thought, one immediately is faced with its
unpredictable nature, which brings up the question: how the seeming
unpredictability of thought could be reconciled with the stable excitation
patterns approach? This question could be answered in more than one way, and
different mechanisms of thought generation could actually be present. \
Thought generation could be related to (sub) patterns rearrangement and
generating new patterns from subpatterns of different patterns, so that new
patterns could be accepted and made permanent if they ''pass certain tests''
(equivalent to ''are able to excite certain other patterns'' in the
P-machine language). In addition, the networks' excitation patterns are
modified by many inputs, thus varying the patterns excited at any given time
by the same network. These patterns are also excitation modes of the
networks responsible for the direction of learning and evolution of thought.
At the same time, there is a great deal of stability of such patterns:
similar thoughts come to us, as well as similar emotional patterns are
excited, in response to similar situations. Also, there seems to be a
continuous learning process by the network which results in the excitation
patterns being continually modified. This process could also be described by
the excitation of patterns, our patterns of learning and patterns (or ways)
of thinking appear to be stable over our lifetimes.

Let me stress again that the described language and thought generating
hierarchy is presented here as a possibility and to illustrate the
theoretical framework suggested in this paper. More research, experimental,
theoretical, and computer modeling - based, is needed to understand
mechanisms of human language and thought.

\section{Generalizations}

One of the points of this paper is that an intelligent system could be
formed by hierarchies of P-machines consisting of pattern recognition and
pattern excitation networks which could be trained though Hebbian\ learning
process mediated by a local binary interaction. The question arises:\ do
there exist systems other than nervous, which are capable of forming PRNs
and PENs with Hebbian-like local learning rules? Such systems could in
principal exist at different temporal and spatial scales and might not be
easily detectable in the human lifetime. Human societies, from small groups
to civilizations, and ecosystems could be viewed from such a point of view.
Are these examples just parts of a greater hierarchy which at a certain
level also includes the human brain? In other words, what other non-linear
interactions in nature, apart from the ones mediated by chemical
neurotransmitters, could be responsible for formation of intelligent systems
at different temporal and spatial scales?

\section{Discussion and open questions}

\begin{enumerate}
\item  Experimental status. What would be the experimental strategy to
observe P-machines of the brain? Since neuron itself is a P-machine, the
question applies to the macroscopic P-machine containing large number of
neurons.

Pattern recognition and excitation networks in the brain might be difficult
to isolate. These networks could be overlapped, interleaved and entangled in
the brain tissues. Moreover, at least in principle, same neurons can
participate in many different circuits by switching inputs on and off using
different synapses, which could significantly increase the number of
excitation (and recognition) patterns attainable with a given number of
neurons. This could created difficulties for direct experimental observation
of such networks. Still the following questions arise:\ Can the networks
described in chapter \ref{ch example language} or their analogs be actually
identified in the human brain? Do there exist training delay networks
connecting, for example, sound recognition with the sound excitation
circuits in the brain which could be observed experimentally?

Other evidence could be based on injuries and ablations. For example,
ablations in Wernicke's and Broka's areas result in different types of
speech and thought impairments. Does it shed any light on the neural
P-machine hierarchy organization in the brain apart from the mere location
of the impaired functions?

Our numerous experiences tell us that we enjoy certain repeating activities
and patterns. Does the reason for this lie in the (resonance) excitation of
patterns of activity in the brain?\ Examples of such resonance excitations
would be music, poetry, dancing and other activities with their
characteristic repeating sounds and rhythms. Do the consonant and dissonant
chords which are characterized by sound frequency ratios equal to certain
prime numbers open a window into the workings of the brain? Can excitation
modes of some of the brain's neural networks be characterized by ratios of
prime numbers in the ways similar to the mathematical description of linear
resonatorls, among them sound modes of strings and membranes?

\item  d-networks play an important role in translating recognized patterns
into the training sets for excitation networks, not just passive delay
elements synchronizing pattern recognition - excitation training processes.
Details of this translation are not clear at this time and need further
investigation.

\item  The notion of scale plays the central role in modern science. It gave
rise to such ideas in theoretical physics and mathematics as renormalization
and wavelet analysis. What is the analog of scale in the world of patterns?
One possible answer to this question is: the number of hierarchical levels
in the pattern, which could be considered its complexity. The structure of
the Fig. \ref{ladder} is capable of recognizing complex patterns on its
inputs and generating complex responses to them. It scales in the sense that
patterns of higher complexity can be handled with correspondingly larger
number of levels consisting of blocks of Fig. \ref{basic-learning}. One can
make one more step in determining the brain's coarse grained degrees of
freedom and try these structures for the role of the building blocks of the
brain.

\item  Two papers considering how people play games were recently published 
\cite{chess1}, \cite{chess2}, which further substantiate the idea that games
are played by humans as pattern excitations in response to patterns
recognized in positions, not by mere calculating next moves. This is
different from the traditional Artificial Intelligence (AI) approach in
which games are played based on the analysis of the position tree arising
from different moves. Which approach is ultimately more powerful in
situations where computing or network resources are limited is an open
question, however the latest chess games between the human world champion
and computers show that computers seem to be winning. In my view this could
be attributed to the limit of number of specific game patterns the human
player comes across during his playing experience, and therefore the number
of recognizable game patterns and excitation modes which could be trained
over the human lifetime.

\item  Emotions represent excitation patterns which have the ability to
modify the work of other pattern recognition and generating networks.
Subconsciousness could be viewed as the set of excitation patterns, which
are excited in response to their stimuli, but which people are not
necessarily aware of. These excitations, being modes of non-linear
resonators, could give rise to or block other excitations in various parts
of the system, leading to various, sometimes undesired consequences \cite
{freud},\cite{freud ego and Id}. From the point of view of this paper,
psycho-analysis is a way to re-train the blocking processing pathways, or
bypass them altogether by creating new ones.

\item  The number of recognized and excitable patterns in living organisms
is limited by the available resources of their nervous systems. Given the
average time scale of learning\ to recognize and excite a pattern, there are
only limited number of patterns an organism can potentially learn and excite
over its lifetime. The need in the number of patterns is also limited by the
lifetime needs of the organism. As new conditions arise in the environment,
organisms learn to recognize them and develop new excitation patterns to
generate responses. Discovery of new patterns in nature which are important
for living organisms is a very laborious process. In the process of
evolution new levels of the hierarchy are added to the neural networks to
make organisms more adaptable to various environmental challenges.

\item  Ideas discussed in this paper could be applied to various fields such
as understanding and treatment of mental and neurological diseases,
education, artificial intelligence systems and so on. However, for practical
applications, many properties of the pattern excitation neural networks and
their training will have to be studied in great details. These studies must
involve both identification of the pattern excitation networks in nervous
systems, analytical investigation and computer simulations of such networks.
\end{enumerate}

\noindent \textbf{Acknowledgment}\textsl{\ }\textit{I would like to thank
A.Portnoy and K.Wickman for their encouragement and support.}

\pagebreak

\end{document}